\documentclass[epj,nopacs]{svjour}

\usepackage{latexsym}
\usepackage{graphics}
\usepackage{balance}

\journalname{}
\begin{document}
\sloppy
\title{An ultracold, optically trapped mixture of $^{87}$Rb and metastable $^{4}$He atoms}
\titlerunning{An ultracold, optically trapped mixture of $^{87}$Rb and metastable $^{4}$He atoms}
\author{Adonis Silva Flores\thanks{e-mail: \texttt{a.s.flores@vu.nl}} \and Hari Prasad Mishra \and Wim Vassen \and Steven Knoop\thanks{present address: Royal Netherlands Meteorological Institute (KNMI), De Bilt, The Netherlands; e-mail: \texttt{steven.knoop@knmi.nl}}}
\institute{LaserLaB, Department of Physics and Astronomy, VU University, De Boelelaan 1081, 1081 HV Amsterdam, The Netherlands}
\date{\today}

\abstract{
We report on the realization of an ultracold ($<25~\mu$K) mixture of rubidium ($^{87}$Rb) and metastable triplet helium ($^{4}$He) in an optical dipole trap. Our scheme involves laser cooling in a dual-species magneto-optical trap, simultaneous MW- and RF-induced forced evaporative cooling in a quadrupole magnetic trap, and transfer to a single-beam optical dipole trap. We observe long trapping lifetimes for the doubly spin-stretched spin-state mixture and measure much shorter lifetimes for other spin-state combinations. We discuss prospects for realizing quantum degenerate mixtures of alkali-metal and metastable helium atoms.
}

\maketitle
\section{Introduction}\label{intro}

Ultracold mixtures of distinct atomic species serve for many scientific goals: sympathetic cooling of atomic species for which evaporative cooling is inefficient \cite{hadzibabic2002tsm,modugno2001bec,truscott2001oof}, creation of ultracold polar molecules \cite{ni2008ahp}, exploring many-body physics in quantum degenerate Bose-Bose, Bose-Fermi and Fermi-Fermi atomic mixtures \cite{ferrierbarbut2014amo,yao2016ooc,roy2016tem}, studying impurities immersed in Bose or Fermi gases \cite{spethmann2012dos,scelle2013mco,hu2016bpi,cetina2015doi}, heteronuclear few-body physics \cite{pires2014ooe,tung2014oog}, and testing universality of free-fall \cite{schlippert2014qto}.

Most mixtures of chemically distinct atomic species consist of alkali-metals: Li+Na \cite{hadzibabic2002tsm}, Li+K \cite{yao2016ooc,spiegelhalder2010aop,wu2011sii}, Li+K+Rb \cite{taglieber2008qdt}, Li+Rb \cite{silber2005qdm}, Li+Cs \cite{mosk2001scw,tung2013uma,repp2013ooi}, Na+K \cite{park2012qdb}, Na+Rb \cite{wang2016narbbec}, K+Rb \cite{modugno2001bec,roati2002fbq,wacker2015tds}, K+Cs \cite{grobner2016anq}, Rb+Cs \cite{lercher2011poa,mccarron2011dsb}. These experimental efforts laid the foundation of the creation of ultracold heteronuclear ground-state molecules, which posses a large permanent dipole moment: KRb \cite{ni2008ahp}, RbCs \cite{takekoshi2014uds,molony2014cou}, NaK \cite{park2015udg} and NaRb \cite{guo2016coa}. Another example is the observation of successive Efimov states in Li+Cs \cite{pires2014ooe,tung2014oog}, benefiting from the largest possible mass ratio within the alkali-metal group. In most of these cases preparation in an optical dipole trap \cite{grimm2000odt} is essential, in particular to allow for tunable interaction and magneto-association by means of magnetically induced Feshbach resonances \cite{chin2010fri}. 

Recently also mixtures of alkali-metal and Yb or Sr atoms have become available: Rb+Yb  \cite{nemitz2009poh,vaidya2015dbf},
Li+Yb \cite{roy2016tem,hara2011qdm,hansen2011qdm}, Rb+Sr \cite{pasquiou2013qdm}, and efforts towards Cs+Yb \cite{kemp2016pac}. Here the main interest comes from the doublet $^2{\rm \Sigma}^+$ molecular ground state potential that gives rise to both electric and magnetic tunability of the associated molecules, in contrast to the singlet $^1{\rm \Sigma}^+$ ground state potential of bialkali-metal molecules. However, for these systems only very narrow Feshbach resonances are expected  \cite{zuchowski2010urm,brue2012mtf,brue203pof}, and so far no resonance has been observed. This has triggered work towards mixtures involving metastable Yb($^3P_2$) \cite{khramov2014uhm,dowd2015mfd}, for which broader resonances are expected, however accompanied with strong inelastic two-body losses \cite{gonzalezmartinez2013mtf,petrov2015mco}.

Here we have realized an optically trapped, ultracold mixture of an alkali-metal and helium in the metastable 2~$^3$S$_1$ state (He$^*$). Ultracold mixtures of alkali-metal atoms and fermionic $^3$He$^*$ or bosonic $^4$He$^*$ provide new Bose-Bose, Bose-Fermi and Fermi-Fermi mixtures, with an extended range of possible mass ratios. The scattering properties of He$^*$+alkali-metal collisions are described by a shallow quartet $^4{\rm \Sigma}^+$ potential and a deeper doublet $^2{\rm \Sigma}^+$ potential. Accurate \emph{ab initio} calculations of the quartet $^4{\rm \Sigma}^+$ potentials and the corresponding quartet scattering lengths have recently become available for Li, K, Na and Rb \cite{kedziera2015aii}, and while most of the doublet $^2{\rm \Sigma}^+$ potentials have been studied experimentally and theoretically in the past (see e.\,g.\,\cite{cohen1985tco,ruf1987tio,scheibner1987roa,merz1990eat,movre2000tio}), the doublet scattering lengths are unknown. Importantly, the large internal energy of 19.8~eV of He$^*$ leads to Penning ionization (PI):
\begin{equation}
{\rm He}(2 ^3{\rm S}_1)+{\rm A}(^2{\rm S}_{1/2}) \rightarrow {\rm He}(1 ^1{\rm S}_0)+{\rm A}^+(^1{\rm S}_0)+{\rm e}^-\label{PI},
\end{equation}
resulting in trap loss. Fortunately, PI is suppressed for pure quartet scattering due to spin-conservation, which is essential for realizing a stable ultracold mixture. This requires He$^*$ and the alkali-metal atom to be both prepared in either the low-field or high-field spin-stretched spin-state. For other spin-state combinations, relevant for Feshbach resonances, the PI loss rate depends on the amount of doublet character of the particular entrance channel. A crucial assumption here is that the first excited (non-spin-singlet) A$^+$ state is energetically not available, which is true for all alkali-metal atoms except Cs. 

\begin{figure}
\center
\resizebox{0.35\textwidth}{!}{%
\includegraphics{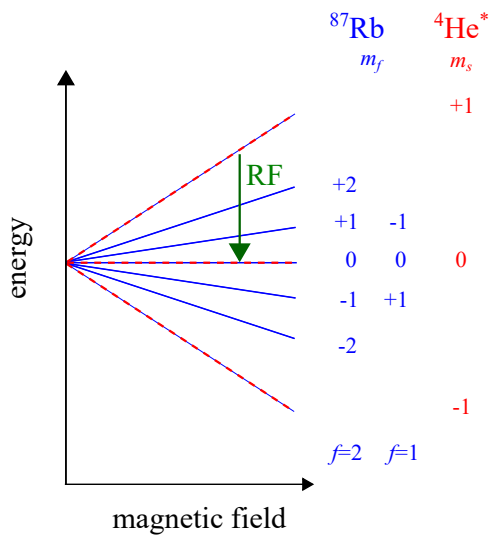}
}
\caption{(Color online) Ground state Zeeman splitting diagram of $^{87}$Rb (blue) and $^4$He$^*$ (red-dashed) showing the possible different spin-state combinations that can be prepared. Also depicted in the diagram is the difference in the Zeeman splitting between He$^*$ (2$\mu_{\rm B}B$) and Rb ($\mu_{\rm B}B$/2) allowing the possibility to transfer He$^*$ atoms between the Zeeman states using RF without affecting Rb. The hyperfine splitting between the $f=2$ and $f=1$ states of $^{87}$Rb is 6.835~GHz and a MW sweep is used to transfer Rb atoms between these states.
\label{HeRbZeeman}}
\center
\end{figure}

Our experiment involves an ultracold mixture of $^4$He$^*$+$^{87}$Rb, for which dual-species laser-cooling and trapping was first achieved by the Truscott group \cite{byron2010tli}. Magnetic trapping of the stable, doubly spin-stretched spin-state combination $|m_s=+1\rangle_{^{4}{\rm He}^*}+|f=2,m_f=+2\rangle_{^{87}{\rm Rb}}$ (see Fig. \ref{HeRbZeeman}) has also been reported \cite{byron2010sop,knoop2014umo}, providing upper limits of the PI loss rate for quartet scattering on the order of $10^{-12}$~cm$^3$s$^{-1}$, and revealing a small quartet scattering length in agreement with \emph{ab initio} calculations \cite{knoop2014umo,kedziera2015aii}. 

We have recently reported on measurements of the two-body loss rate coefficients for different spin-mixtures in the optical dipole trap, which are compared with predictions of multichannel quantum-defect theory \cite{flores2016qsc}. In this article, we focus on the experimental realization of the ultracold mixture in the optical dipole trap, giving a detailed discussion on the different preparation stages and emphasize the challenges of the mixture compared to the single-species situation. Finally, we present lifetime measurements, comparing the doubly spin-stretched spin-state with the energetically lowest spin-state combination $|m_s=-1\rangle_{^{4}{\rm He}^*}+|f=1,m_f=+1\rangle_{^{87}{\rm Rb}}$, and outline the consequences for dual-species quantum degeneracy and Feshbach spectroscopy.

\section{Experiment}\label{exp}

\begin{figure}
\center
\resizebox{0.45\textwidth}{!}{%
\includegraphics{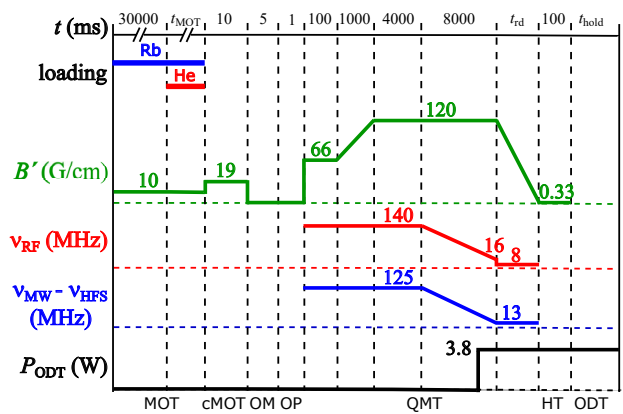}
}
\caption{(Color online) Summary of the loading scheme for the preparation of the ultracold mixture of $^{87}$Rb and $^4$He$^*$ in an optical trap. Shown are the magnetic field gradient, MW and RF frequencies, and ODT power corresponding to the different trapping stages. Typical MOT loading time $t_{\rm MOT}$ of He$^*$ is less than 1~s and the QMT ramp down duration $t_{\rm rd}$ is between 0.5 to 2~s.
\label{loadingschememix}}
\center
\end{figure}

Details of our experimental setup have been given in Ref. \cite{knoop2014umo} wherein we describe our interspecies thermalization measurement in a magnetic trap, Ref. \cite{mishra2015epo}, in which we implement a hybrid trap (single-beam optical dipole trap combined with a weak quadrupole magnetic trap) to achieve Bose-Einstein condensation of  $^{87}$Rb  and Ref. \cite{flores2015smf}, where we demonstrate the production of $^4$He$^*$ Bose-Einstein condensates in a single-beam optical dipole trap. In this paper, we focus on the scheme that is relevant for simultaneous loading of the two atomic species in an optical dipole trap. We start by loading both species in a three-dimensional magneto-optical trap (3D-MOT), after which the mixture is further cooled in an optical molasses (OM). After optical pumping (OP) to the desired low-field seeking doubly spin-stretched spin-state, the sample is transferred to a quadrupole magnetic trap (QMT) for further evaporative cooling using microwave (MW) and radiofrequency (RF) for Rb and He$^*$, respectively. Finally, the sample is loaded into a single-beam optical dipole trap (ODT) using a hybrid trap (HT) as an intermediate stage. A summary of the loading scheme starting from the MOT stage towards the ODT is illustrated in Fig. \ref{loadingschememix}.

In the following subsections, we briefly describe the different stages involved in the preparation of our ultracold mixture and discuss important issues regarding simultaneous loading as compared to our single-species experiments \cite{mishra2015epo,flores2015smf} and previous mixture experiment \cite{knoop2014umo}. We also describe our simultaneous detection scheme and explain the preparation of different spin-state samples using MW and RF frequency sweeps. An overview of the Zeeman states is given in Fig. \ref{HeRbZeeman}.

\subsection{Two-species MOT}\label{MOT}

Our 3D-MOT consists of three 2-inch retro-reflected beams containing the cooling light of both species. Changing from 1-inch \cite{knoop2014umo,mishra2015epo} to 2-inch 3D-MOT beams was motivated by the fact that sympathetic cooling of He$^*$ with Rb is not efficient, therefore forced evaporative cooling is required and thus the need to start with sufficient He$^*$ atom number. An order of magnitude improvement is observed in the He$^*$ 3D-MOT atom number as already reported in Ref. \cite{flores2015smf}. Correspondingly, the Rb 3D-MOT atom number also improves by at least a factor of three compared to Ref. \cite{mishra2015epo}.

The quadrupole magnetic field is derived from a pair of coils in anti-Helmholtz configuration, and the magnetic field gradient is 10~G/cm along the weak axis. The total 3D-MOT incoming power for Rb is around 20~mW and the light is detuned by -20~MHz with respect to the laser cooling transition (natural linewidth is 6~MHz). For He$^{*}$, the light is detuned by -33~MHz, about 20 times the natural linewidth of 1.6~MHz, to reduce light-assisted intraspecies Penning ionization loss \cite{rooijakkers1997lda,mastwijk1998oco,tol1999lno}. The total incoming 3D-MOT power is around 30~mW. We deliver these beams to the setup via polarization maintaining (PM) fibers where we couple the 3D-MOT beams of the two species in the same fiber using dichroic mirrors.
Rb atoms are loaded from a 2D-MOT with two distinct pushing beams \cite{knoop2014umo,park2012cab} and He$^{*}$ atoms are loaded from a Zeeman slower. We first load the Rb atoms in 30~s followed by He$^{*}$ loading typically in less than 1~s. To minimize the continuous flux of ground state He (metastable to ground state fraction is 10$^{-4}$), an in-vacuo shutter is introduced between the source and the Zeeman slower, which is only open during the He$^{*}$ loading. Typically, we lose between 15 to 20\% of the Rb atoms in the 3D-MOT due to the combined effect of background collisions with thermal ground-state and metastable He atoms. At the end of this stage, we have at least 3$\times10^9$ (Rb) and 3$\times10^8$ (He$^*$) atoms at a temperature of a few hundreds of $\mu$K (Rb) and 1 to 2~mK (He$^*$), respectively. After loading the two atomic species, we compress the sample by abruptly increasing the gradient to 19~G/cm while the detunings are ramped to $-15~{\rm MHz}$ (Rb) and $-5~{\rm MHz}$ (He$^*$) in 10~ms (cMOT stage). Afterward, the 3D-MOT gradient is switched off and the power of the 3D-MOT beams is reduced by almost a factor of 10 for further cooling in the OM stage. However, the power imbalance between the incoming and reflecting 3D-MOT beams limits the allowed duration of the OM stage in order for the clouds not to deviate too far from the center of the QMT. To compensate for this, we offset the alignment of the 3D-MOT beams with respect to the QMT center such that at the end of the OM, the clouds coincide with the position of the QMT. After a 5~ms OM stage, we apply simultaneous OP (around 150~$\mu$W each) on both species for a duration of 1~ms in order to prepare the sample in the doubly spin-stretched spin-state before transferring to the QMT for further cooling.

\subsection{Simultaneous evaporative cooling in the QMT}\label{evapQMT}

For magnetic trapping, we have used the same coils to create the quadrupole magnetic trap as used in the 3D-MOT. After the OP stage, we abruptly increase the gradient to 66~G/cm and stay for 100~ms in order to facilitate the initial transfer to the QMT. The gradient is then ramped up to 120~G/cm in 1~s. We typically transfer about 30 to 40\% of the atoms of both species from the 3D-MOT into the magnetic trap. Since sympathetic cooling is not efficient due to the small interspecies scattering length and large mass ratio \cite{kedziera2015aii,knoop2014umo}, we perform simultaneous MW-induced (Rb) and RF-induced (He$^{*}$) forced evaporative cooling in the QMT. To generate the MW frequencies, we use a 6.8~GHz phase-locked oscillator (Amplus PLO) mixed with the frequency doubled output of a tunable 80~MHz signal generator. After a series of filtering and amplification stages, we send between 1 to 2~W of power to a MW horn. For the RF, we also use a tunable 80~MHz signal generator that is frequency doubled. We send around 5~W to an RF coil after a series of amplification stages.

\begin{figure}
\center
\resizebox{0.45\textwidth}{!}{%
\includegraphics{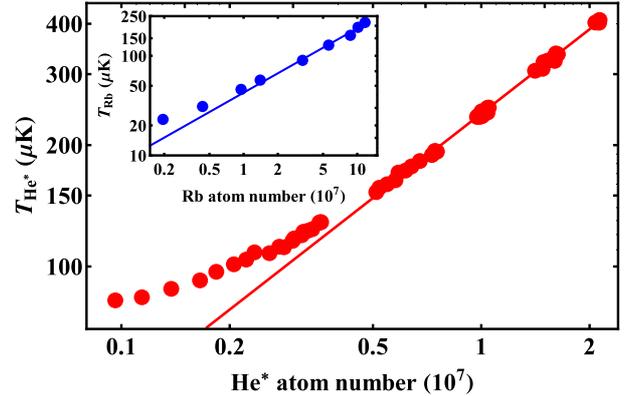}
}
\caption{(Color online) He$^*$ (red circles) and Rb (inset) temperatures as function of atom number during the evaporative cooling in the QMT. Solid lines are double-logarithmic fits that include only the first few points indicating the efficiency of the evaporation process. Data are measured in a single-species experiment.
\label{HeRbQMT}}
\center
\end{figure}

\begin{figure}
\center
\resizebox{0.45\textwidth}{!}{%
\includegraphics{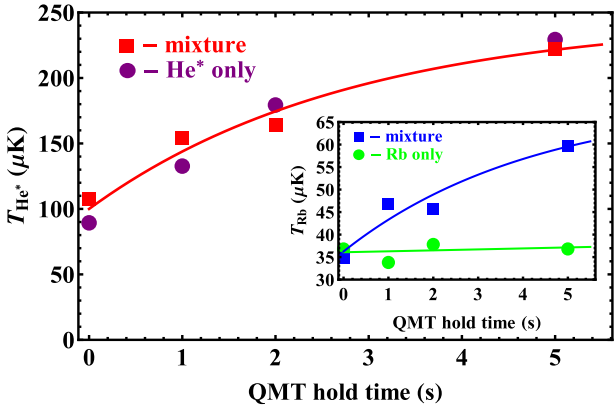}
}
\caption{(Color online) Time evolution of He$^*$ and Rb (inset) temperatures after the final stage of the evaporative cooling in the QMT. Solid lines are a guide to the eye. The increasing He$^*$ temperature is due to the Majorana heating. In the case of Rb, wherein the Majorana heating is still negligible (at least an order of magnitude slower than He$^*$ as exhibited in the green data), the temperature increase (blue data) is due to thermalization with the higher temperature He$^*$. 
\label{QMTHeRbtempevol}}
\center
\end{figure}

The lowest temperature that can be achieved by evaporative cooling in a QMT is limited by Majorana loss and heating. The Majorana effect scales inversely with mass \cite{petrich1995stc,dubessy2012rbe} and is more pronounced for light atomic species such as He$^*$. In Fig. \ref{HeRbQMT}, we show the measured temperature and number of atoms of He$^*$ and Rb during the evaporation process in the QMT. It is clear that the efficiency of the evaporative cooling for He$^*$ starts to go down at a much higher temperature (150~$\mu$K) as compared to the heavier Rb (below 50~$\mu$K). The data suggest that the lowest temperature that can be achieved for He$^*$, while maintaining sufficient number of atoms, will be higher than for Rb. This temperature difference has to be taken into account during the simultaneous evaporative cooling because the overall efficiency of the evaporation process in the QMT will be affected by the interplay between the Majorana heating and interspecies thermalization. Generally, this means that we want to keep the temperatures of the two species as close as possible to minimize heating of Rb due to interspecies thermalization with the hotter He$^*$ atoms. An example of such a scenario is shown in Fig. \ref{QMTHeRbtempevol}. The increasing trend in the He$^*$ temperature is due to Majorana heating. There is no significant difference in the He$^*$ temperature with or without the presence of the Rb atoms. This is because Majorana heating dominates over the small effect of interspecies thermalization with Rb atoms. On the other hand, for the colder Rb sample wherein the Majorana effect is still small, the interspecies thermalization with hotter He$^*$ atoms dominates and causes a noticeable increase in the Rb temperature.

Other important issues to be considered during the simultaneous evaporation process are the MW and RF frequencies. The MW (between 6.8~GHz to 7.0~GHz) used for evaporative cooling of Rb will not affect the He$^{*}$ atoms in the magnetic trap, but the RF (160~MHz and lower) used for He$^{*}$ in principle can be in resonance with the Rb atoms. However, the Zeeman splitting of $^4$He$^*$ is larger than that of $^{87}$Rb (see Fig. \ref{HeRbZeeman}), which is a general feature in He$^*$+alkali-metal mixtures. For Rb in the $|f=2,m_f=+2\rangle$ state, the trap depth given by the RF is a factor of two larger than that for He$^*$, and therefore RF can be selectively used for He$^*$. The condition that the MW-truncated trap depth of Rb is lower than the RF-truncated one is given by: $\nu_{\rm MW}-\nu_{\rm HFS}<3\nu_{\rm RF}$, where $\nu_{\rm RF}$ and $\nu_{\rm MW}$ are the RF and MW frequencies respectively, and $\nu_{\rm HFS}$ is the hyperfine splitting of Rb. For the two species to have equal trap depths, the condition is $\nu_{\rm MW}-\nu_{\rm HFS}=3\nu_{\rm RF}/2$.

During the initial transfer to the QMT, we send a fixed $\nu_{\rm MW}-\nu_{\rm HFS}$=125~MHz and $\nu_{\rm RF}$=140~MHz. After 4~s of thermalization and plain evaporation, we initiate forced evaporative cooling of the mixture. We use a total evaporation time of 8~s, in which we ramp down $\nu_{\rm MW}-\nu_{\rm HFS}$ to 13~MHz and $\nu_{\rm RF}$ to 16~MHz. These values correspond to trap depths of 430~$\mu$K (Rb) and 800~$\mu$K (He$^*$). At the end of the simultaneous evaporative cooling, we can have a few 10$^6$ atoms for both Rb and He$^*$ at temperatures of around 50~$\mu$K and 90~$\mu$K, respectively.

\subsection{Transfer to single-beam ODT}\label{transferODT}

\begin{figure}
\center
\resizebox{0.45\textwidth}{!}{%
\includegraphics{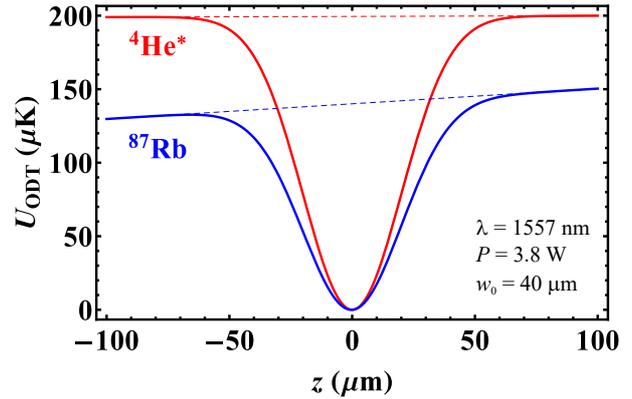}
}
\caption{(Color online) ODT trapping potentials along the radial (gravity) direction of He$^*$ (red) and Rb (blue). Note the asymmetry of the Rb potential due to gravity and the higher trap depth of He$^*$ due to the higher polarizability.
\label{ODTHeRb}}
\center
\end{figure}

Our ODT light has a wavelength $\lambda=1557$~nm. A piezo-controlled mirror is used to precisely align the ODT beam with respect to the QMT center \cite{mishra2015epo,flores2015smf}. The ratio in the polarizability of He$^*$ and Rb is 1.4 \cite{safronova2006fdp,notermans2014mwf}. With the available ODT power of around 4~W and a waist of 40~$\mu$m, the corresponding trap depths are around 200~$\mu$K (He$^*$) and 140~$\mu$K (Rb), respectively. In Fig. \ref{ODTHeRb}, we show the ODT potentials of $^4$He$^*$ and $^{87}$Rb for an ODT power of 3.8~W. The effect of gravity is noticeable for Rb by the asymmetry of the trapping potential, leading to a slight reduction of the trap depth. At low ODT powers, the differential gravitational sag will lead to a separation of the two clouds. Here, we stay at a high ODT power at which this effect is negligible.

\begin{figure}
\center
\resizebox{0.45\textwidth}{!}{%
\includegraphics{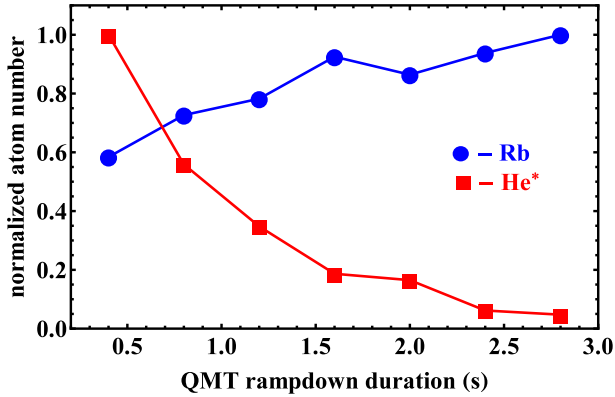}
}
\caption{(Color online) Normalized atom number showing the loading of He$^*$ (red) and Rb (blue) into the ODT as a function of the QMT ramp down duration. Solid lines are a guide to the eye.
\label{RbandHeloadingvsQMTramp}}
\center
\end{figure}

After evaporative cooling in the QMT, the mixture is transferred to a hybrid trap (HT) by ramping down the gradient of the QMT to the levitation gradient (0.33~G/cm) of $^4$He$^*$ \cite{flores2015smf}. Here, we only use the  HT as a bridging stage to facilitate the transfer to the pure ODT. Loading the mixture into the HT or ODT is not as straightforward as in our single-species experiment \cite{mishra2015epo,flores2015smf}. The difference in the initial conditions, such as temperature and density, and the different properties such as mass and polarizability imply different loading conditions. Most of these parameters are coupled and difficult to disentangle and investigate individually. Here, we focus on parameters that are crucial in the simultaneous loading and can be tuned in the loading scheme. In Fig. \ref{RbandHeloadingvsQMTramp}, we show the number of atoms loaded in the ODT as a function of the duration of QMT gradient ramp down, comparing Rb and He$^*$. There is a clear difference in the duration for optimum loading between the two atomic species. Ideally, loading should be slow enough (adiabatic transfer), such that the atoms can smoothly follow the transition from the QMT to the ODT potential, as in the case of Rb, where the optimum transfer is toward longer duration. This is not surprising because we load Rb at a temperature that is much lower than the ODT trap depth. In fact, from our single-species experiment, we already observe a saturation in the loading of Rb above 2.5~W \cite{mishra2015epo}. In the case of He$^*$, the optimum transfer appears to be at a shorter duration of the QMT ramp down. A plausible explanation is because the temperature (limited by the Majorana effect) at which we load He$^*$ is just about half of the ODT trap depth. Correspondingly, from the single-species experiment we do not see a clear sign of saturation in the loading \cite{flores2015smf}. Furthermore, during the QMT ramp down, the He$^*$ cloud expands much faster than Rb due to the higher temperature and smaller mass. Here, we hypothesize that the geometrical size of the cloud with respect to the geometry of the ODT potential has a mismatch, limiting to short duration for the QMT ramp down in the case of He$^*$. On an absolute scale, we generally observe that the transfer efficiency for Rb is higher than for He$^*$. In future experiments, this can be improved by using higher ODT powers at which the He$^*$ loading can be also saturated. 

\begin{figure}
\center
\resizebox{0.45\textwidth}{!}{%
\includegraphics{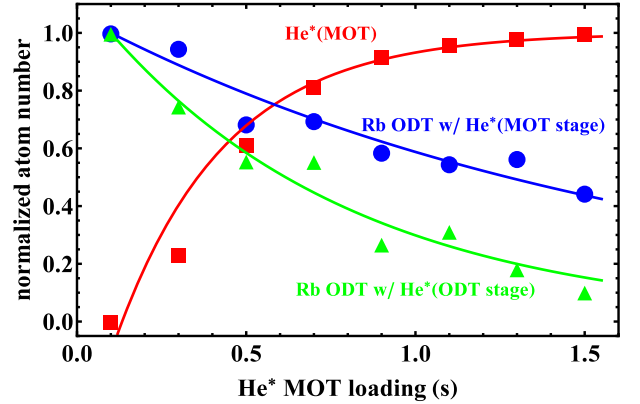}
}
\caption{(Color online) Rb atoms loaded in the ODT as a function of He$^*$ MOT loading time, comparing two situations. First is with He$^*$ atoms loaded up to the MOT stage (blue circles). Second is with He$^*$ atoms loaded up to the ODT stage (green triangles). The corresponding He$^*$ atoms during MOT loading is also shown (red squares). Solid lines are a guide to the eye.
\label{RbODTvsHeload}}
\center
\end{figure}

More general, aside from the small interspecies loss in the MOT stage, there are no additional losses for He$^*$ as long the mixture stays in the doubly spin-stretched spin-state. For Rb, this is not the case particularly during the evaporative cooling in the QMT wherein the He$^*$ atoms can be a heat load for the Rb atoms as already discussed in Sect. \ref{evapQMT}. Additionally, the flux of ground state He that made it to the main chamber during the MOT loading also introduces additional loss for the Rb atoms. These issues are summarized in Fig. \ref{RbODTvsHeload}, where we plot the number of Rb atoms loaded in the ODT as a function of He$^*$ MOT loading time. For the blue circles, the He$^*$ atoms are only introduced up to the MOT stage. Out of the total 50\% loss in the number of Rb atoms in the ODT, around 20\% can be accounted from the MOT stage (for a fully loaded He$^*$) while the remaining 30\% can be explained due to background collisions during the QMT stage and transfer to the ODT. For the green triangles, the He$^*$ atoms are present up to the ODT stage. The additional loss in the number of Rb atoms is due to heating from the hotter He$^*$ atoms during the simultaneous evaporative cooling in the QMT. Together with the QMT ramp down duration, we use the He$^*$ MOT loading to tune the ratio in the atom numbers of He$^*$ and Rb in the ODT. In our mixture experiment, we typically load He$^*$ in the 3D-MOT between 0.4~s to 0.8~s while the QMT ramp down duration is between 0.5~s to 2~s. We can tune between (0.2-1)$\times10^5$ atoms for Rb and He$^*$ at temperatures around 15~$\mu$K (Rb) and 22~$\mu$K (He$^*$).

\subsection{Dual absorption imaging and MCP detection}\label{detection}

Standard absorption imaging is used to measure the atom number and temperature of the samples. To save optical access in our main chamber, we also couple the imaging beams of the two species in the same fiber using a dichroic mirror. To create the proper circular polarization, we implement a 920~nm zero-order quarter waveplate. We use a dichroic mirror to image the two clouds onto two different cameras. For Rb, we use a CCD camera (QImaging Exi-blue) with 6.45~$\mu$m pixel size. For He, we use an InGaAs camera (Xenics Xeva 320) with a 30~$\mu$m pixel size. We use a 2:1 (L1 = 30~cm and L2 = 15~cm) imaging to accommodate the size of the cloud onto the camera chip. Absorption imaging of the two species can be done simultaneously. This is essential, especially during the optimization process, in which we can easily track the positions of the two clouds when aligning the 3D-MOT beams to optimize the OM stage and the subsequent transfer of the mixture to the QMT.

Additionally for He$^*$, we also use a microchannel plate (MCP) detector that is positioned below the trap center (at angle 22$^\circ$ with respect to the direction of gravity) to measure the time-of-flight (TOF) distribution. A magnetic gradient pulse from a single deflection coil is applied to direct the atoms onto the MCP detector \cite{flores2015smf}. TOF signals of He$^*$ can also be obtained simultaneously with Rb imaging. After the clouds are released from the ODT, we first capture the images using ballistic expansions typically between 1 to 5~ms after which we apply the magnetic gradient pulse.

Our experiments involve mixtures of different spin-state combinations, for which state-selective detection is essential. In the case of our MCP TOF detection, the orientation of the deflection coil and the MCP position will already suffice for the purpose. In principle, we can only detect He$^*$ atoms in the $|m_s=+1\rangle$ state. To detect atoms in the $|m_s=-1\rangle$ state, we apply an RF sweep that transfers back the atoms to the $|m_s=+1\rangle$ state. For Rb, we use the repumping light during imaging to distinguish atoms between the two hyperfine states $|f=2,m_f=+2\rangle$ and $|f=1,m_f=+1\rangle$.

\subsection{Preparation of different spin-states}\label{spinstates}

To prepare a mixture of different spin-state combinations in the optical dipole trap, we perform rapid adiabatic transfer around a magnetic field of 2.5~G. For He$^*$, a 0.5~MHz RF sweep is applied to transfer the atoms between the Zeeman states. A 25~ms RF sweep of around 1~W transfers all of the atoms from the $|m_s=+1\rangle$ to $|m_s=-1\rangle$ state. We confirm the transfer by applying a second sweep that transfers back the atoms to the $|m_s=+1\rangle$. For Rb, a 70~ms MW sweep of 0.2~MHz is used to transfer the atoms from the hyperfine state $|f=2,m_f=+2\rangle$ to $|f=1,m_f=+1\rangle$. However, we only manage to transfer 50\% of the atoms to the $|f=1,m_f=+1\rangle$, due to limited MW power. We immediately send resonant light for 15~ms to clean the remaining atoms in the $|f=2,m_f=+2\rangle$. Among the various possible combinations, we work with $|m_s=-1\rangle+|f=2,m_f=+2\rangle$ (single RF sweep on He$^*$), $|m_s=+1\rangle+|f=1,m_f=+1\rangle$ (single MW sweep on Rb) and $|m_s=-1\rangle+|f=1,m_f=+1\rangle$ (MW sweep on Rb followed by RF sweep on He$^*$). For the given magnetic field, the RF sweep used for He$^*$ does not affect the Rb atoms following a similar argument as described earlier in Sect. \ref{evapQMT}.

\begin{figure}
\center
\resizebox{0.45\textwidth}{!}{%
\includegraphics{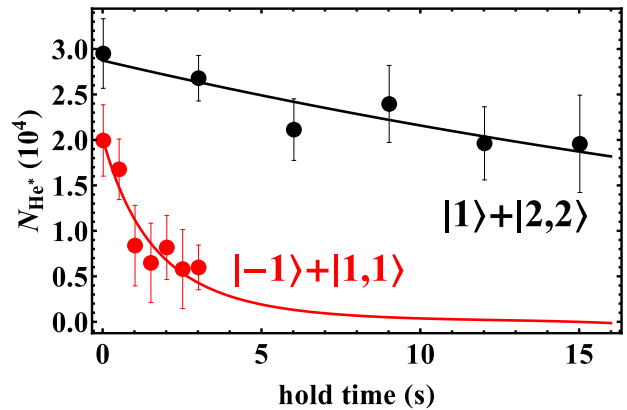}
}
\caption{(Color online) Remaining He$^*$ atoms in the ODT as a function of hold time showing the stable spin-state combination $|m_s=+1\rangle+|f=2,m_f=+2\rangle$ (black) and the shorter lifetime spin-state combination $|m_s=-1\rangle+|f=1,m_f=+1\rangle$ (red). The initial Rb atom numbers are $N_{\rm Rb}=(8.6\pm0.7)\times 10^4$ for $|m_s=+1\rangle+|f=2,m_f=+2\rangle$ and $N_{\rm Rb}=(4.5\pm0.8)\times 10^4$ for $|m_s=-1\rangle+|f=1,m_f=+1\rangle$. The approximately factor two difference in the initial Rb number is due to the MW transfer efficiency. Solid lines are numerical fit from the solution of Eq. \ref{PIloss}, from which the two-body loss rates are obtained.
\label{mixturelifetimedoubly}}
\center
\end{figure}

\section{Lifetimes in optical dipole trap}\label{lifetimes}

For measurements of trapping lifetimes, we ramp down the QMT gradient in 2~s to allow for a smooth (and adiabatic) transfer. Afterward, we hold the mixture in the ODT for 2.5~s to ensure thermalization of each species before preparing a particular spin-state combination. We need to measure the initial temperatures of both species because the interspecies thermalization rate (0.01~$\rm s^{-1}$) is absent on the experimentally relevant time scales. In here, we can assume a constant temperature of each species during lifetime measurements. We hold the mixture in the ODT for a variable time and measure the remaining atoms in the trap. We can measure both the remaining Rb and He$^*$ atoms simultaneously from the absorption imaging (Rb) and MCP TOF detection (He$^*$). However, the signal-to-noise ratio from MCP detection is better than from absorption imaging. In this regard, our analysis of trap loss is based mostly on He$^*$ MCP data and we only use the Rb data as a counter-check \cite{flores2016qsc}.

We measure a long trapping lifetime for the doubly spin-stretched spin-state $|m_s=+1\rangle+|f=2,m_f=+2\rangle$ mixture (see Fig. \ref{mixturelifetimedoubly}, black circles). We observe short lifetimes of a second to a few seconds for the other spin-state combinations, for which Penning ionization is spin-allowed. An example is also shown in Fig. \ref{mixturelifetimedoubly} (red circles) for the case of $|m_s=-1\rangle+|f=1,m_f=+1\rangle$.

\begin{figure}
\center
\resizebox{0.45\textwidth}{!}{%
\includegraphics{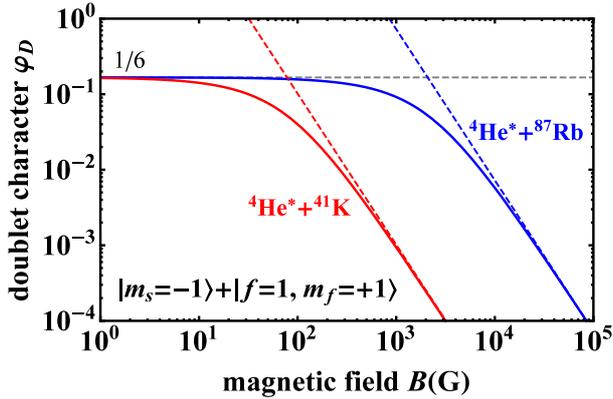}
}
\caption{(Color online) Doublet character $\varphi_D$ as function of magnetic field for the energetically lowest spin-state combination $|m_s=-1\rangle+|f=1,m_f=+1\rangle$, comparing $^4$He$^*$+$^{87}$Rb (blue) and $^4$He$^*$+$^{41}$K (red). Horizontal dashed line corresponds to $\varphi_D=1/6$ and the dashed colored lines give the asymptotic $B^{-2}$ dependence.
\label{RbvsK}}
\center
\end{figure}

To extract the interspecies two-body loss rates $L_2$ of the different spin-state combinations, we fit the trap loss data with the numerical solution of the two coupled equations,
\begin{equation}\label{PIloss}
\frac{d}{dt}N_{\rm i}=-{\rm \Gamma}_{\rm i} N_{\rm i}-L_2 \int n_{\rm i}(\vec{r})n_{\rm j}(\vec{r})d\vec{r},
\end{equation}
where i or j can be assigned interchangeably to He$^*$ or Rb, $N$ is the atom number, ${\rm \Gamma}$ is the one-body loss rate, $n(\vec{r})=n_0\exp[-U(\vec{r})/k_BT]$ is the density, $n_0=N/\int\exp[-U(\vec{r})/k_BT]d\vec{r}$ is the peak density and $U(\vec{r})$ is the trapping potential wherein the contribution due to gravity is included. The one-body loss rates ${\rm \Gamma}$ are measured independently. Intraspecies two- and three-body loss processes, including homonuclear Penning ionization, can be fully neglected for the chosen spin-states under our conditions \cite{soding1999tbd,marte2002fri,borbely2012mfd}.

For the loss rate in the stable $|m_s=+1\rangle+|f=2,m_f=+2\rangle$ mixture, we do not observe a significant difference compared to the single-species measurement, which suggests that it is only limited by background collisions. We obtain an upper limit in the total two-body loss rate, $L_2=1.3\times10^{-12}{\rm cm}^3{\rm s}^{-1}$, similar as previous upper limits obtained from magnetic trap experiments \cite{byron2010sop,knoop2014umo}.

For the other spin-state combinations, we obtain relatively large $L_2$ coefficients on the order of $10^{-11}-10^{-10}{\rm cm}^3{\rm s}^{-1}$ \cite{flores2016qsc}. For the energetically lowest spin-state combination $|m_s=-1\rangle+|f=1,m_f=+1\rangle$, for which Penning ionization is the only possible two-body loss process, we obtain $L_2=5.3^{+2.0}_{-1.7}\times10^{-11}{\rm cm}^3{\rm s}^{-1}$. This value is not far from the universal rate constant $L_2=4.5\times10^{-10}{\rm cm}^3{\rm s}^{-1}$ \cite{idziaszek2010urc}, if one takes into account an additional factor that represent the amount of doublet character ($\varphi_D$), of the particular spin-state combination, which at small magnetic fields is 1/6 in this case. This observation suggests that at high magnetic fields, where for this spin-state combination $\varphi_D$ goes to zero, two-body loss will be reduced. The magnetic field dependence of $\varphi_D$ is given in Fig. \ref{RbvsK}, showing the case of $^{87}$Rb and $^{41}$K. The transition from a constant $\varphi_D=1/6$ to a decreasing $\varphi_D\sim B^{-2}$ occurs at a magnetic field that scales with the hyperfine splitting. Thus, while this effect occurs for $^{87}$Rb at rather high magnetic fields, for $^{41}$K, which has the lowest hyperfine splitting of 254~MHz, this effect can be observed at experimentally feasible magnetic fields.

\section{Conclusions and prospects}\label{concl}

In conclusion, we have realized an ultracold, optically trapped mixture of $^{87}$Rb and $^4$He$^*$ atoms. We have demonstrated simultaneous RF-induced (He$^*$) and MW-induced (Rb) forced evaporative cooling in the quadrupole magnetic trap. We have measured a long trapping lifetime for the doubly spin-stretched spin-state combination exhibiting a strong suppression of Penning ionization loss. Realizing a dual-species BEC is in principle possible using this spin-state combination. Here, it is crucial to compensate (or control) the differential gravitational sag that leads to separation of the two clouds due to the huge mass ratio between Rb and He$^*$. Among the possible solutions are the application of special optical dipole trap geometries that allow for strong vertical confinement in a shallow trap (see e.\ g.\ \cite{yao2016ooc}), or the addition of an optical dipole beam that selectively supports the heavy species \cite{ulmanis2016utb}. Another approach is to add a magnetic field gradient that will provide an additional species-dependent force \cite{hansen2013poq}.

Realizing a BEC of $^{87}$Rb or $^4$He$^*$ in a single-species preparation is relatively straightforward using our existing hybrid trap or single-beam ODT scheme \cite{mishra2015epo,flores2015smf}. However for the mixture preparation, constraints as outlined in this paper limit the transfer efficiency (and thus the initial phase-space density: at least an order of magnitude smaller compared to our single-species preparation) into the hybrid trap or single-beam ODT. The main issue is the Majorana effect that limits the lowest achievable temperature for He$^*$ in the magnetic trap. This is not a fundamental limit but rather an experimental consequence of using a QMT, and not present in an Ioffe-Pritchard type of magnetic trap. Another approach is to use higher ODT powers that provide higher trap depths, such that it is not necessary to push the evaporative cooling towards lower temperature in the QMT \cite{bouton2015fpo}. For the case of Rb and He$^*$, we estimate that an ODT power of around 10~W will already provide sufficient transfer (a few $10^6$ atoms at temperatures below 30~$\mu$K) in the HT or single-beam ODT for evaporative cooling towards dual quantum degeneracy.

Using an analogous experimental scheme as described in this paper (i.e. simultaneous MW (alkali-metal) and RF (He*) evaporative cooling), ultracold bosonic $^4$He$^*$ and bosonic alkali-metal mixtures all seem possible. For the fermionic alkali-metal atoms, the cooling strategy will rely on sympathetic cooling with $^4$He$^*$ and thus depend on a favorable interspecies quartet scattering length \cite{kedziera2015aii}. Similarly, for fermionic $^3$He$^*$ and bosonic alkali-metal mixtures, realizing two-species quantum degeneracy also depends on the quartet scattering length. In here, $^4$He$^*$ can also be introduced to sympathetically cool $^3$He$^*$ \cite{mcnamara2006dgb}. Finally, for $^3$He$^*$ and fermionic alkali-metal mixtures, sympathetic cooling with a third species is required which can either be $^4$He$^*$ or another bosonic alkali with favorable quartet scattering length.

Feshbach resonances are in principle possible due to the hyperfine coupling between the doublet and quartet interaction potentials. This requires a mixture in a spin-state combination other than the purely quartet doubly spin-stretched spin-state and is thus accompanied by strong two-body loss, which limits the scattering length tunability around the Feshbach resonance \cite{hutson2007fri} and the observation of (enhanced) three-body recombination loss. Still, Feshbach spectroscopy can be performed, as we expect a modification of the Penning ionization loss rate around the Feshbach resonances due to coupling between the doublet and quartet interaction potentials.

\begin{acknowledgement}
We acknowledge Rob Kortekaas for excellent technical support. This work was financially supported by the Netherlands Organization for Scientific Research (NWO) via a VIDI grant (680-47-511) and the Dutch Foundation for Fundamental Research on Matter (FOM) via a Projectruimte grant (11PR2905). 
\end{acknowledgement}

\bibliographystyle{phpf}
\balance
\bibliography{HeRbLib}

\end{document}